\documentclass[journal]{vgtc}

\usepackage{graphicx}
\DeclareGraphicsExtensions{.eps}
\graphicspath{{figures/}{pictures/}{images/}{./}}

\usepackage{microtype}
\PassOptionsToPackage{warn}{textcomp}
\usepackage{textcomp}
\usepackage{mathptmx}
\usepackage{times}

\usepackage{cite}
\usepackage{tabu}
\usepackage{booktabs}

\usepackage{datetime}
\vgtcinsertpkg 

\title{Cyclic Polygon Plots}

\author{Maksim Schreck\thanks{e-mail: maksim.schreck@iwr.uni-heidelberg.de}
\and Peter Albers\thanks{e-mail: peter.albers@uni-heidelberg.de} %
\and Filip Sadlo\thanks{e-mail: sadlo@uni-heidelberg.de}}
\affiliation{\scriptsize Heidelberg University, Germany}

\abstract{In this paper, we introduce the cyclic polygon plot, a representation based on a novel projection concept for multi-dimensional values. Cyclic polygon plots combine the typically competing requirements of quantitativeness, image-space efficiency, and readability. Our approach is complemented with a placement strategy based on its intrinsic features, resulting in a dimensionality reduction strategy that is consistent with our overall concept. As a result, our approach combines advantages from dimensionality reduction techniques and quantitative plots, supporting a wide range of tasks in multi-dimensional data analysis. We examine and discuss the overall properties of our approach, and demonstrate its utility with a user study and selected examples.%
}


\teaser{
  \centering
  \newcommand{\mywidth}{0.2355\linewidth}%
	\subfloat[\label{fig:teaser-pcp}]{%
	  \includegraphics[width=\mywidth]{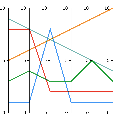}%
	}%
        \hfill%
	\subfloat[\label{fig:teaser-rc}]{%
	  \includegraphics[width=\mywidth]{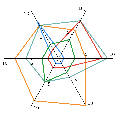}%
	}%
        \hfill%
	\subfloat[\label{fig:teaser-cpp-abbc}]{%
	  \includegraphics[width=\mywidth]{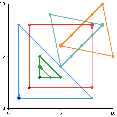}%
	}%
        \hfill%
	\subfloat[\label{fig:teaser-cpp-abcd}]{%
	  \includegraphics[width=\mywidth]{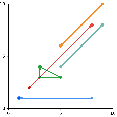}%
	}%
	\caption{\label{fig:teaser} Synthetic dataset, featuring a pulse~(blue), step-down (red), linearly ascending (orange), linearly descending (cyan), and double peak (green) sequence.
          Parallel coordinates plot~\protect\subref{fig:teaser-pcp} and radar chart~\protect\subref{fig:teaser-rc}, for comparison with our cyclic polygon plot with \abbc~\protect\subref{fig:teaser-cpp-abbc} and \abcd~\protect\subref{fig:teaser-cpp-abcd} cyclic pair selection.
          Notice ``start arrows'' and darker color indicating multiple instances in \protect\subref{fig:teaser-cpp-abbc} and \protect\subref{fig:teaser-cpp-abcd}.
\vspace*{6pt}
}}

\usepackage{xcolor}
\usepackage{tikz}
\usepackage[outline]{contour}
\contourlength{1pt}
\contournumber{20}

\usepackage{amssymb}						
\usepackage{amsmath}
\usepackage{mathtools}
\usepackage{gensymb}

\renewcommand{\vec}[1]{\ensuremath{\mathbf{#1}}} 

\newcommand{\eqsep}{\,}
\newcommand{\twoeqsep}{\quad}
\newcommand{\R}{\mathbb{R}}
\newcommand{\Rn}{\mathbb{R}^n}
\newcommand{\Mod}[1]{\ (\mathrm{mod}\ #1)}

\newcommand{\abbc}{\text{\mbox{ab-bc}}}
\newcommand{\abcd}{\text{\mbox{ab-cd}}}
\newcommand{\datavec}{\vec{d}}
\newcommand{\dataveccomp}{\delta}
\newcommand{\vertex}{\vec{v}}

\newcommand{\ji}{\theta}
\newcommand{\Sc}{\tau}

\usepackage{subfig}
\usepackage{caption}
\usepackage[export]{adjustbox}

\newcommand{\threeImgWidth}{0.326\linewidth}

\newcommand{\fourImgWidth}{0.243\linewidth}

\newcommand{\sixImgWidth}{0.159\linewidth}

\usepackage[percent]{overpic}
\usepackage{pstricks}

\usepackage[noabbrev,capitalise]{cleveref}

\crefdefaultlabelformat{#2#1#3}
\creflabelformat{equation}{#2#1#3}

\crefname{equation}{Equation}{Equations}
\crefname{figure}{Figure}{Figures}
\crefname{tabular}{Table}{Tables}
\crefname{section}{Section}{Sections}

\Crefname{equation}{Eq.}{Eqs.}
\Crefname{figure}{Fig.}{Figs.}
\Crefname{tabular}{Tab.}{Tabs.}
\Crefname{section}{Sec.}{Secs.}

\usepackage{dblfloatfix}
\usepackage{tabularx}
\usepackage{multirow}
\usepackage{booktabs}

\usepackage{enumitem}

\begin{document}

\firstsection{Introduction}
\label{sec:introduction}

\maketitle

There is hardly a real-world question that could be answered by considering a single quantity. In fact, many considerations require mutual analysis of a large number of attributes, necessitating effective means for multi-dimensional data analysis.

A wide field of visualization techniques has been proposed for analyzing such multi-dimensional data, and ultimately, all of them need to perform some kind of projection from the multi-dimensional data domain to 2D image space. For example, projections employed in dimensionality reduction techniques typically involve continuous distortion that, on the one hand, aims to reduce clutter by decoupling visual density from data dimension, and on the other hand, tries to preserve original properties, such as distance metrics. Although very successful in various fields, the involved projection and distortion of the \emph{data} cause loss of information and loss of quantitativeness, i.e., the original data cannot be determined from dimensionality reduction results. Other techniques, such as the parallel coordinates plot~(PCP) and radar chart~(RC), avoid such losses by employing discrete projections, which project the \emph{axes} of the multi-dimensional data domain to 2D image space, and use these projected axes to represent the data. However, despite being quantitative and information-preserving, these techniques do typically not scale well with higher data dimensions, due to inferior image-space utilization and involved readability issues.

Overall, combining the typically competing requirements of quantitativeness, image-space efficiency, and readability is a main challenge in multi-dimensional data visualization. With the cyclic polygon plot, we present an approach that combines these requirements. It shares quantitative readability of PCPs and RCs, and the comparably high image-space efficiency of scatterplots~(SP). We achieve this by splitting the data domain into two-dimensional subspaces and projecting these subspaces to image space with superposition. Thus, a multi-dimensional value represents a point in each of these 2D subspaces, and accordingly a set of points in their superposition in image space, which we connect to a polygon to preserve the correspondence to the original dimensions. In that regard, cyclic polygon plots represent a generalization of scatterplots to multi-dimensional data, generalizing points to polygons. Due to symmetry considerations, we choose the subspaces from the data domain in a cyclic manner, motivating the name of the resulting approach. Two variants of this subspace choice proved useful, exhibiting slightly different advantages, and denoted \abbc{} and \abcd{} scheme. We demonstrate that our polygons also serve well as glyphs, in particular when placed according to their intrinsic properties, providing a consistent dimensionality reduction approach.

The contributions of this work include:
\begin{itemize}[noitemsep,topsep=-6pt]
\item cyclic polygon plots,
\item placement of cyclic polygon plots, and
\item their detailed discussion and evaluation including a user study.
\end{itemize}

\section{Related Work}
\label{sec:related-work}

Spatialization (positioning of values in a parameterized space) of data-vector values for creating polygons is widely discussed, most prominently in the context of star glyphs~\cite{fienberg1979Graphical} and RCs~\cite{draper2009Survey}. Whereas the glyph placement is often not intrinsically defined (as is the case with star glyphs), and often performed with a grid layout or with first and second data-vector attributes as spatialization dimensions~\cite{ward2002Taxonomy}, the placement of glyphs is also discussed in a geographical context~\cite{opach2018Star}. Fuchs et al.~\cite{fuchs2013Evaluation} discuss the viability of different glyph designs while leveraging the small multiples principle, reinforced by advantageous glyph placement. In our method, we provide an intrinsic mapping of our glyphs to spatial position, and discuss additional, geometrically motivated placement strategies.
Radial axes layouts, as present in, for example, star glyphs, RCs, RadViz~\cite{hoffman1997DNA}, or star coordinates~\cite{kandogan2000Star}, feature a compact and intuitive way to represent data~\cite{burch2014benefits,tominski2004Axesbased}. However, they tend to aggravate analysis, since they are harder and less efficient to interpret~\cite{waldner2020Comparison,goldberg2011Eye}. With our approach, we provide a line- and polygon-based visualization, which does not rely on radial axes but is embedded in a 2D space, which is more familiar to interpret~\cite{chan2006Survey}.

The PCP~\cite{inselberg1985plane} is a widely used and expanded multivariate visualization technique. Its extension to 3D has been discussed in various contexts~\cite{falkman2001Information,streit20063D} and configurations~\cite{johansson20053dimensional,wegenkittl1997Visualizing}. Different axis layouts exist, most notably the use of a common attribute across all dimensions~\cite{falkman2001Information} and a bipartite layout of axes~\cite{johansson2014usability}. Often, an extension to 3D is employed to reduce cluttering in high-density areas~\cite{ahmed2006GEOMI}, but almost always signifies the need for user interactivity to benefit from the 3D visualization layout. 
When interpreting our cyclic polygon approach as a projection of a 3D-PCP (see below), we use a different and more intuitive axis layout which will be discussed in more detail in \cref{sec:method}. Fanea et al.~\cite{fanea2005interactive} present a different method to extend the PCP to 3D by integrating it with star glyphs. Similar to our approach, this approach also supports a frontal and lateral projection, which in this case results in a star glyph and PCP representation, respectively.
Zhou et al.~\cite{zhou2018IndexedPoints} introduce an indexed point representation of generalized $p$-dimensional flat surfaces from $n$-variate data. The resulting indexed points are represented in a lower-variate PCP. Analogous to our cyclic polygons, this approach also performs a mapping of $n$-variate data to, in this case, 3D subspaces.
Claessen et al.~\cite{claessen2011flexible} provide a framework for interactive design of a representation consisting of multiple, arbitrarily placed coordinate axes with PCPs or scatterplots displayed between them.
While the interactive design promotes data exploration, it also requires domain knowledge to fully leverage its flexibility. With our approach, we try to limit this prior knowledge and provide a self-sufficient data representation.
Blaas et al.~\cite{blaas2008extensions} implement a GPU-based processing pipeline to effectively display large datasets using PCPs. One of the main pipeline tasks identified by them is normalization. 

Nam and Mueller~\cite{nam2013TripAdvisor} use a trip metaphor to introduce an iterative and interactive visualization approach. This results in an overview map consisting of glyphs where the user can control parameters to navigate and enlarge the visualization. Contrary to our approach, one glyph represents a single subspace of the data in contrast to all subspaces of the data, as it is done in our approach.
The subspace voyager~\cite{wang2018Subspace} is an extension to the previous approach. Here, explicit mapping to 3D subspaces is combined with an integrated navigation interface which aims to improve usability concerning manual exploration.

The representations of scatterplot matrix~(SPLOM)~\cite{cleveland1988dynamic} and generalized plot matrix~\cite{im2013gplom} can also be understood as 2D subspace mappings of $n$-variate input data, whereas the parallel scatterplot matrix~\cite{viau2010flowvizmenu} additionally provides a detail view of selected dimension pairs. Nevertheless, none of these approaches combine the subspaces into a single \textit{common} 2D space.

\section{Method}
\label{sec:method}

There are two variants that our approach naturally leads to, the \abbc{} and the \abcd{} scheme. We first motivate the overall approach (\cref{sec:motivation}), followed by a description of the \abbc{} scheme (\crefrange{sec:cyclic-pair-selection}{sec:mapping}). Subsequently, we describe the minor modification that gives rise to the \abcd{} scheme (\cref{sec:scheme-abcd}), and provide a detailed discussion of the properties of both schemes (\cref{sec:properties}). Finally, as a complementing approach, we investigate the suitability of our polygons as glyphs in placement strategies derived from the polygons themselves and their properties (\cref{sec:placement}).

\subsection{Motivation}
\label{sec:motivation}

Our design is motivated by the aim of combining the multi-dimensional quantitativeness of, e.g., parallel coordinates plots with the image-space efficiency of scatterplots and dimensionality reduction techniques, while maintaining readability.
We realize these requirements in a novel approach that utilizes subspace mapping to represent $n$D data in a single 2D image space, while maintaining correspondence to the data dimensions by representing them as a polygon.
Alternatively, it can be interpreted as 
a generalization of scatterplots from 2-dimensional (bivariate) to $n$-dimensional (multivariate) data, while keeping the representation two-dimensional.
Overall, we ultimately need to map the $n$-dimensional data domain to two-dimensional image space in a quantitative manner.

\subsection{Cyclic Pair Selection}
\label{sec:cyclic-pair-selection}

We start with decomposing the $n$-dimensional data domain into a sequence of $k$ two-dimensional subspaces. That is, each $n$-dimensional value%
\begin{equation}
  \datavec \coloneqq (\dataveccomp_0,\dots,\dataveccomp_{n-1})
  \in \Rn
\end{equation}
is transformed into a sequence of $k$ 2D vertex (subspace) coordinates
\begin{equation}
  \vertex_j \coloneqq \left(x_j,y_j \right) \in \R^2 \eqsep, \twoeqsep 0 \leq j \leq k-1 \eqsep.
\end{equation}

We achieve this by what we call \emph{cyclic pair selection}. We denote the most fundamental variant of such selection the \emph{\abbc{} scheme}:
\begin{equation}
  \label{eq:cyclic-pair-selection-abbc}
  \datavec = (\dataveccomp_0,\dots,\dataveccomp_{n-1}) \mapsto ( \dataveccomp_j,\dataveccomp_{j+1 \Mod{n}})_{j=0,\dots,n-1} \eqsep .
\end{equation}
Here, we iterate through the $n$-dimensional value and sequentially pick adjacent pairs of components as one vertex coordinate. In other words,%
\begin{equation}
  \datavec = (\dataveccomp_0,\dots,\dataveccomp_{n-1}) \mapsto (\dataveccomp_0,\dataveccomp_1),(\dataveccomp_1,\dataveccomp_2),\dots,(\dataveccomp_{n-1},\dataveccomp_0)
  \eqsep .
\end{equation}
We regard this scheme fundamental, because, due to its sequential overlap and
cyclic closure, it does not cause unnecessary loss of generality, i.e., it is invariant to cyclic permutation of $\datavec$. That is,
\begin{equation} 
  \tilde{\datavec} \coloneqq (\dataveccomp_{l \Mod{n}},\dots,\dataveccomp_{n-1+l \Mod{n}})
\end{equation}
produces the same sequence of 2D vertices, simply shifted by $l$. The reason for this invariance is that the \abbc{} selection scheme is order-preserving and maps each $\dataveccomp_i$ equally to the $x$- and $y$-coordinate of the 2D vertices, i.e., it does not induce bias. For the \abbc{} scheme,  $k=n$, i.e., it decomposes the $n$-dimensional data domain into a sequence of $n$ two-dimensional subspaces.

To not exceed the scope of our work, we assume the order of attributes in the $n$-dimensional value to be invariant.
However, optimization of this ordering is possible and has been widely discussed in the context of PCPs~\cite{lu2016Two,lu2020DoubleArc} and generally in multi-dimensional visualization~\cite{peng2004Clutter,artero2006Enhanced,hahsler2008Getting,johansson2009Interactivea,dasgupta2010Pargnostics}.
For space reasons, we include noteworthy details and relations about these permutations and resulting geometric variation in the supplemental material.

\subsection{Mapping}
\label{sec:mapping}

The obtained 2D subspaces satisfy our requirement of being quantitative. Therefore, the second step for our transformation from the $n$D data domain to 2D image space is to map and integrate these 2D subspaces to a single image space.

Here comes image-space utilization into play.
One design strategy could be to place the 2D subspaces in matrix arrangement in image space, which would directly lead to the superdiagonal of the scatterplot matrix. Scatterplot matrices, however, tend to waste image space with redundant display of subspaces including their axes.
In addition, these issues also affect readability.
Parallel coordinates plots, as well as radar charts, share some of these shortcomings regarding waste of image space and readability (see also \cref{sec:results-userstudy}). Beyond that, their axes and ticks tend to clutter with the polyline content. Furthermore, due to their point--line duality with scatterplots, they tend to suffer from additional clutter, because points in the original 2D subspaces are mapped to entire line segments. Due to our cyclic pair selection, however, our technique resides in the point domain of the point--line duality (as discussed below), and thus tends to reduce such clutter.

Overall, these observations lead to the following requirements:
\begin{itemize}[noitemsep,topsep=-6pt]
\item avoid side-by-side placement of the 2D subspaces,
\item avoid visual representation of more than two axes, and
\item keep the axes and their ticks outside of the content area.
\vspace*{6pt}
\end{itemize}

\begin{figure}[t]%
  \centering%
  \subfloat[\label{fig:polygon-creation-illustration-combined}]{%
    \begin{overpic}[height=0.385\linewidth]{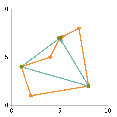}%
      \newcommand{\myfontsize}{0.55}
      \put(58,65.5){\rput[c](0,0){\color{black} \scalebox{\myfontsize}{(\textit{i})}}}%
      \put(49.5,73){\rput[c](0,0){\color{black} \scalebox{\myfontsize}{$(\dataveccomp_0,\dataveccomp_1)$}}}%
      \put(68,81.0){\rput[c](0,0){\color{black} \scalebox{\myfontsize}{$(\dataveccomp_1,\dataveccomp_2)$}}}%
      \put(82,21){\rput[c](0,0){\color{black} \scalebox{\myfontsize}{$(\dataveccomp_2,\dataveccomp_3)$}}}%
      \put(28,14){\rput[c](0,0){\color{black} \scalebox{\myfontsize}{$(\dataveccomp_3,\dataveccomp_4)$}}}%
      \put(18,49){\rput[c](0,0){\color{black} \scalebox{\myfontsize}{$(\dataveccomp_4,\dataveccomp_5)$}}}%
      \put(46.5,46){\rput[c](0,0){\color{black} \scalebox{\myfontsize}{$(\dataveccomp_5,\dataveccomp_0)$}}}%
    \end{overpic}%
  }%
  \hfill%
  \subfloat[\label{fig:3d-pcp}]{%
    \includegraphics[height=0.385\linewidth]{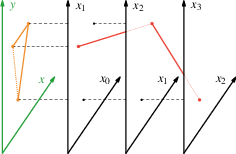}%
  }%
  \caption{\label{fig:3d-pcp-and}
    \protect\subref{fig:polygon-creation-illustration-combined}~Cyclic polygon plot~(CPP) with \abbc{}~(orange) and \abcd{}~(cyan) scheme, including data components~$\dataveccomp_i$.
    \protect\subref{fig:3d-pcp}~CPP~(orange) as projection of modified 3D parallel coordinates plot~\cite{johansson2014usability}~(red).
  }%
\end{figure}

These requirements motivate us to superimpose all 2D subspaces using an identity transformation, i.e., to share the same origin, abscissa, and ordinate.
This merged representation is mapped to image space for display, 
generally using linear, or, if beneficial, logarithmic scaling (\cref{sec:wine-dataset}). As a consequence, all vertices~$\vertex_j$ are mapped to dots in image space, together with a depiction of the abscissa and the ordinate. This transformation, being an identity transformation, does not introduce any distortion to the representation of the displayed data. We remove the individual axes of the subspaces in favor of a single axis pair, a sacrifice necessary to achieve a screen-space efficient layout. Due to our subspace generation, this does not pose a threat to the quantitativeness of our representation. It does, however, require a final step to convey the correspondence between the dots in image space and the components~$\dataveccomp_i$ of the multi-dimensional value~$\datavec$. 
We achieve this by connecting cyclically adjacent vertices of the sequence, i.e., we draw edges $\smash{\overline{\vertex_j\vertex_m}}$, with $m=j+1\Mod{k}$ and $j=0,\dots,k-1$. This results in a polygon (\cref{fig:polygon-creation-illustration-combined}, orange), whose connectivity represents the vertex sequence from the cyclic pair selection (\cref{eq:cyclic-pair-selection-abbc}). For absolute readability of the resulting representation, one needs to additionally be able to identify the first vertex and the order of the vertex sequence, i.e., the orientation of the polygon. We achieve this by placing an arrow symbol at the first vertex~$\vertex_0$ in direction of the second vertex~$\vertex_1$ ((\textit{i}) in \cref{fig:polygon-creation-illustration-combined}). Notice that we set the size of this symbol in the order of the size of the vertex dot to avoid interference in case of dense data. We call the resulting visualization \emph{cyclic polygon plot}~(CPP), and, compared to scatterplots, it draws a polygon for each multi-dimensional data value~$\datavec$, instead of a single dot.
Since the polygons may overlap themselves, as well as other polygons (as discussed below), we use blending in the rendering step to convey such cases (see, e.g., the dark blue dot in \cref{fig:teaser-cpp-abbc}).

\subsection{Alternative Cyclic Pair Selection Scheme: \abcd{}}
\label{sec:scheme-abcd}

Before we come to the discussion of the properties of the CPP, let us have a quick look at an alternative scheme for cyclic pair selection that proved beneficial for reducing visual clutter. We denote this the \emph{\abcd{} scheme}, and define it as (cf.\ \cref{eq:cyclic-pair-selection-abbc}):
\begin{equation}
  \label{eq:cyclic-pair-selection-abcd}
  \datavec = (\dataveccomp_0,\cdots,\dataveccomp_{n-1}) \mapsto (\dataveccomp_{\,2j},\dataveccomp_{\,2j+1 \Mod{n}})_{j=0,\cdots,p-1} \eqsep ,
\end{equation}
with $p\coloneqq \lceil n / 2 \rceil$. In other words, for even $n$,
\begin{equation}
  \datavec = (\dataveccomp_0,\cdots,\dataveccomp_{n-1}) \mapsto (\dataveccomp_0,\dataveccomp_1),(\dataveccomp_2,\dataveccomp_3),\cdots,(\dataveccomp_{n-2},\dataveccomp_{n-1}) \eqsep ,
\end{equation}
and for odd $n$,
\begin{equation}
  \label{eq:cyclic-pair-selection-abcd-odd}
  \datavec = (\dataveccomp_0,\cdots,\dataveccomp_{n-1}) \mapsto (\dataveccomp_0,\dataveccomp_1),(\dataveccomp_2,\dataveccomp_3),\cdots,(\dataveccomp_{n-1},\dataveccomp_{0}) \eqsep .
\end{equation}
That is, in case of odd dimension of the multi-dimensional value~$\datavec$, its first component~$\dataveccomp_0$ is repeated for the last vertex. Overall, for the \abcd{} scheme, $k=p$, i.e., it decomposes the $n$-dimensional data domain into a sequence of $p$ two-dimensional subspaces. Regarding the mapping (\cref{sec:mapping}), the only difference to the \abbc{} scheme is that vertices from \cref{eq:cyclic-pair-selection-abcd} are used instead from \cref{eq:cyclic-pair-selection-abbc}. 

We observe that the \abcd{} scheme is contained in the \abbc{} scheme, i.e., the \abcd{} scheme consists of every second vertex of the \abbc{} scheme (see \cref{fig:polygon-creation-illustration-combined} and \cref{fig:teaser-cpp-abbc,fig:teaser-cpp-abcd}). Nevertheless, due to the sequential overlap of the \abbc{} scheme, this subsampling does not cause loss of information, it simply discards the redundancy contained in the \abbc{} scheme. This reduction has the advantage of reducing the complexity of the visual representation, and thus reducing visual clutter in large datasets, which is our motivation for this scheme.
As is evident from \cref{eq:cyclic-pair-selection-abcd-odd}, the repetition of the first component~$\dataveccomp_0$ in the odd dimension case can introduce bias to the visualization toward this first dimension. However, our study results and application to real datasets show that it has only a negligible effect on the interpretability in practice, and confirm this repetition to be a feasible solution.

\subsection{Properties}
\label{sec:properties}

Let us now investigate some properties of cyclic polygon plots.

\subsubsection{Relation to Previous Work}

As indicated above, our approach represents a generalization of the scatterplot. For $n=2$, both the \abbc{} and \abcd{} schemes result in the traditional scatterplot. For $n>2$, the first vertex of each polygon (of either scheme)
is still identical to the scatterplot of $(\dataveccomp_0,\dataveccomp_1)$.

Our approach can also be interpreted as a projection of a modification of 3D parallel coordinates~\cite{johansson2014usability}, as illustrated in \cref{fig:3d-pcp}. In their work, Johansson et al.\ replace each axis of the traditional PCP by a 2D space spanned by two axes (black arrows in \cref{fig:3d-pcp}). As a consequence, a multi-dimensional value leads to a point in each of their 2D spaces (red dots), which are connected to a polyline (red) in analogy to the traditional PCP. If, in their concept, one replaces their 2D spaces by our 2D subspaces (\cref{eq:cyclic-pair-selection-abbc,eq:cyclic-pair-selection-abcd}), employs orthographic projection (dashed) of the resulting polylines along the ``third'' axis, and closes the resulting polylines (dotted orange), we obtain our CPP (orange) in a common 2D space (green).

\subsubsection{Basic Reading}

A basic task in CPP-based analysis is to determine the original multi-dimensional value~$\datavec$ from a respective polygon. For this, the arrow symbol needs to be identified (\cref{fig:polygon-creation-illustration-combined}). Its coordinates on the abscissa and ordinate (which we also denote $x$- and $y$-coordinates) give us $\dataveccomp_0$ and $\dataveccomp_1$. The direction of the arrow symbol guides us then to the next vertex, whose coordinates are for the \abbc{} scheme $\dataveccomp_1$ and $\dataveccomp_2$, and for the \abcd{} scheme $\dataveccomp_2$ and $\dataveccomp_3$. Assuming that all vertices of a cyclic polygon are distinct (i.e., $\vertex_j \neq \vertex_m \eqsep, \forall j \neq m$), then each dot is shared by exactly two edges. This lets us unambiguously follow the edge to the next vertex, and so on, until we reach the first one, which indicates that we read the full multi-dimensional value~$\datavec$. This reading appears difficult at first sight, but the user study shows that it competes well with reading of PCPs and RCs.

However, the advantage that CPPs are often more image-space efficient and readable than PCPs and RCs, also because they do not need to draw more than two axes and because they keep these axes away from the content, comes at the cost of a weakness: reading becomes more difficult if vertices appear more than once in a sequence, i.e., if $\exists  j \neq m$ such that $\vertex_j = \vertex_m$.

Let us start with the orange polygon in \cref{fig:teaser-cpp-abbc}, an example employing the \abbc{} scheme without multiple vertices. We identify the arrow at coordinates $(5,6)$, followed by $(6,7)$, $(7,8)$, $(8,9)$, $(9,10)$, and $(10,5)$ at the lower right corner of the polygon. This is its last vertex, since the next edge brings us back to the arrow symbol. We identified all six mutually different dots of the six-dimensional value, and we also see that all dots have the same saturation (no one is darker due to blending of multiple dots).

For the red polygon in \cref{fig:teaser-cpp-abbc}, we identify the arrow symbol, and follow via the bottom right dot to the bottom left dot, which is the first dot depicted in darker red, because the blending of its multiple instances resulted in a darker color. Since using brightness as visual variable for ordinal data does not perform well, it would be hard to determine from the color that this dot appears in fact three times. Nevertheless, since it is the only dot that is darker in this polygon, and since there are four distinct dots for a six-dimensional value, one could derive that its multiplicity has to be three. Indeed, almost all configurations with identical dots we investigated, could be determined by graph theory considerations, even if darker color was only used as an indicator that there was more than one dot at the respective location. However, such considerations would be cumbersome and in most applications impede full quantitative reading.

Overall, we draw two conclusions with respect to readability and identical vertices. Firstly, the issue with identical sequence vertices cannot arise if for each multi-dimensional value~$\datavec$, all its components~$\dataveccomp_i$ are distinct. This is no strong requirement in generic cases, since the dimension~$n$ of the value~$\datavec$ is often low. Furthermore, identical values in entire datasets are often considered degenerate, and removed using perturbation or simulation of simplicity~\cite{edelsbrunner2008Reeb}, following the motivation that natural data do not exhibit identical values. Secondly, as we show in our study and demonstrate in our results, CPPs are particularly powerful for qualitative analysis, and lose only a minor part of their quantitativeness if identical vertices are present. Finally, as is illustrated in \cref{fig:teaser-cpp-abbc,fig:teaser-cpp-abcd}, the \abcd{} scheme tends to reduce vertex multiplicity (here, from 4 to 2 for the blue polygon and from 3 to 2 for the red one) due to its overall vertex reduction property.

\subsubsection{Point--Line Duality}

The well-known point--line duality between SPs and PCPs relates a point in the SP to a line segment in the subdomain of the PCP spanned by the respective axis pair. The duality also holds the other way around, i.e., a point in the PCP relates to a line in the SP~\cite{inselberg1985plane}.

Our approach maps, similar to scatterplots, pairs of values to vertices. Therefore, for the \abbc{} scheme, a vertex of the CPP corresponds to a line segment in the PCP. Beyond that, two consecutive vertices of the \abbc{} CPP consist of three consecutive data values, and as such correspond to two consecutive line segments in the PCP. For the \abcd{} scheme, a vertex in the CPP also relates to an edge in the PCP. Two consecutive vertices of the CPP, however, consist of four consecutive data values, and thus correspond to three consecutive line segments in the PCP.

\subsubsection{Slopes and Offsets}
\label{sec:slopes-and-offsets}

Beyond these straightforward relations, we observe interesting and useful relations regarding slopes and offsets. Assuming that all axes in a PCP have the same scaling and offset (such as in \cref{fig:teaser-pcp}) and assuming the distance between the axes is 1, then the slope of a line segment in such a PCP equals the distance of the corresponding vertex in the CPP to the diagonal of the CPP, with slope unit $\sqrt{2}/2$ (in case of equal scaling on both axes). The slope in the PCP is positive if the vertex is above the diagonal in the CPP, and negative if below. For example, all five collinear orange points in \cref{fig:teaser-cpp-abbc} have distance $\sqrt{2}/2$ from the diagonal and are located above it, therefore, the corresponding segments of the PCP have slope $1$, as one can see from the orange polyline in \cref{fig:teaser-pcp}. Analogously, one can see that the cyan line has slope $-1$. Notice that the five intervals of the PCP map to the five collinear vertices in the CPP, and that the sixth vertex in the CPP corresponds to the interval wrapping around from the last PCP axis to its first axis, exhibiting slope $5$ for the cyan polygon. Beyond that, notice that the cyan polygon is shifted along the CPP diagonal toward the origin by one, which corresponds to the values being one unit smaller (see \cref{fig:teaser-pcp}). The above observations hold for both the \abbc{} and the \abcd{} scheme.

There is also a converse relation regarding slopes with the \abbc{} scheme. An edge in the CPP connects two vertices there, and thus relates two consecutive line segments in the PCP, centered at the PCP axis that is shared by the two line segments. Thereby, the slope of the edge in the CPP represents the factor by which the slopes of the two line segments in the PCP differ, i.e., the slope in the CPP is equal to the factor with which one needs to multiply the slope of the left line segment in the PCP to obtain the slope of the line segment to its right.

\subsection{Placement}
\label{sec:placement}

Similar to the PCP and RC, the CPP works well for datasets of medium size, but tends to suffer from overdrawing when applied to larger datasets. This is a drawback inherent to line-based visualization approaches~\cite{nguyen2018DSPCP,rosenbaum2012Progressive}. To alleviate this problem, we reinterpret our CPP polygons as glyphs, scale them down by a factor of $0.05$ (if not stated otherwise) and employ placement, enabling small multiples~\cite{fuchs2013Evaluation}. That is, the position of a CPP polygon is no longer determined by the value of its vertices, but by properties derived from the polygon and mapped to the coordinates of its centroid. This represents a dimensionality reduction technique, which is quantitative and consistent with the glyph it positions. We derive and evaluate (\cref{sec:results-placement}) four different placement strategies.

\subsubsection{Intrinsic Placement}
\label{sec:intrinsic-placement}

We denote this strategy intrinsic, because for the \abcd{} scheme, it does not move the individual polygons. Instead, each polygon is simply downscaled, while fixing the position of its centroid. The centroids of the polygons of the \abbc{} scheme are, however, all located on the diagonal of the CPP, due to the discussed properties of \cref{eq:cyclic-pair-selection-abbc}. Therefore, we translate each polygon from the \abbc{} scheme to the centroid of the corresponding \abcd{} polygon prior to downscaling. \cref{fig:iris-placement-avg} shows intrinsic placement for the glyph of the \abcd{} scheme at the example of the Iris dataset.

\subsubsection{Geometric Placement}
\label{sec:geometric-placement}

Our experience with the CPP, as well as its intrinsic glyph placement, indicated that the polygons are an effective means for qualitative multi-dimensional visualization. 
This motivated us to derive a placement strategy based on the shape of the polygons, i.e., to derive from a CPP polygon two quantities that could define the new $x$- and $y$-coordinate of its centroid. We chose the quite straightforward measures area and circumference, respectively, which performed surprisingly well.

While the computation of the circumference of a polygon is unambiguous and straightforward, different approaches exist to define the area of possibly self-intersecting polygons~\cite{shor1992Detecting}. Firstly, it can be interpreted as the ``footprint'' of the polygon, representing the entire area encased by the polygon, and disregarding inner edges. Secondly, it can be interpreted as the difference between front-facing and back-facing segments of the polygon, this time respecting the intersection of polygon edges. We chose the latter variant, because it more significantly captures the geometry of the polygon and is more continuous w.r.t.\ its variation. It can be calculated using the Gaussian area formula~\cite{boland2000Polygon}, which, for the \abbc{} scheme, can be abbreviated and calculated directly from the high-dimensional value~$\datavec$ according to
\begin{equation}
  \mu_{\abbc} = \frac{1}{2} \left| \sum_{j=0}^{n-1} \left( \delta_j \delta_{j+1\Mod{n}}-\delta_j^2 \right) \right|  \eqsep .
\end{equation}
The area for the \abcd{} scheme can be obtained efficiently by
\begin{equation}
  \mu_{\abcd} = \frac{1}{2} \left| \sum_{j=0}^{p-1} \left( x_j y_{j\Mod {p}}- y_j x_{j\Mod {p} } \right) \right| \eqsep ,
\end{equation}
with $p\coloneqq \lceil n / 2 \rceil$ (as above), and $\vertex_j = (x_j,y_j)$ (\cref{eq:cyclic-pair-selection-abcd}).
\cref{fig:iris-placement-geo} shows an example of geometric placement for the \abcd{} glyph of the Iris dataset. Notice that because this dataset is four-dimensional, the \abcd{} polygons are line segments, which do not possess area. Nevertheless, the resulting placement still
results in a valid representation in this example (\cref{fig:iris-placement-geo-abbc}).

\subsubsection{Angular Placement}
\label{sec:angular-placement}

As a complementary approach to geometric placement which considers the vertex positions, we propose angular placement, which is derived from the signed angles at the polygon vertices. More precisely, the sum of the counter-clockwise angles is mapped to the $x$-coordinate of a polygon's centroid, whereas the sum of the clockwise angles is mapped to its $y$-axis. We employ the \abcd{} scheme for glyph placement due to its clutter-reducing property. Consequently, we do not show angular placement for the Iris dataset, because the \abcd{} scheme results in line-type polygons (\cref{sec:geometric-placement}) due to its four-dimensional data. Therefore, we refer here to the Billiard dataset for an example of angular placement (\cref{fig:billiard-placement-ang}).

\subsubsection{Statistical Placement}
\label{sec:statistical-placement}

Finally, and mainly for comparison, we consider a fourth placement, which sets the $x$-coordinate of the centroid to the mean of all $\dataveccomp_i$ of data value~$\datavec$, and its $y$-coordinate to their standard deviation. Again for the Iris dataset, \cref{fig:iris-placement-stat} shows a respective example.

\section{Results}
\label{sec:results}

Our evaluation is organized into three parts to adequately cover the features of the our approach. We compare to existing techniques and assess advantages and drawbacks with respect to quantitative analysis, readability, feature extraction, and cluttering. 
First, we evaluate basic properties regarding information extraction using selected datasets (\cref{sec:results-plots}). We then move on to evaluating placement (\cref{sec:results-placement}), where we discuss and evaluate our strategies in the context of dimensionality reduction techniques, and provide a quantitative assessment of our clustering. Finally, we confirm these previously discussed results with a user study covering frequent information visualization tasks (\cref{sec:results-userstudy}).

\subsection{Plots}
\label{sec:results-plots}

\begin{figure}[t]
	\centering%
	\subfloat[\label{fig:qcm-pcp}]{%
		\includegraphics[width=\fourImgWidth]{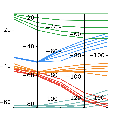}%
	}%
	\hfill%
	\subfloat[\label{fig:qcm-rc}]{%
		\includegraphics[width=\fourImgWidth]{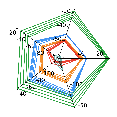}%
	}%
	\hfill%
	\subfloat[\label{fig:qcm-abbc}]{%
		\includegraphics[width=\fourImgWidth]{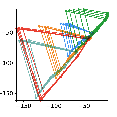}%
	}%
	\hfill%
	\subfloat[\label{fig:qcm-abcd}]{%
		\includegraphics[width=\fourImgWidth]{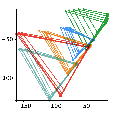}%
	}%
	\caption{\label{fig:qcm}
          5D QCM10 dataset.
          PCP~\protect\subref{fig:qcm-pcp}, RC~\protect\subref{fig:qcm-rc}, CPP with \abbc{}~\protect\subref{fig:qcm-abbc}, and \abcd{}~\protect\subref{fig:qcm-abcd} scheme. Observe linear trend in \protect\subref{fig:qcm-abbc}, \protect\subref{fig:qcm-abcd}.
        }%
\end{figure}

In order to provide a representative overview over the visual properties of our approach,
we examine applications of the CPP ranging from smaller datasets, where features of distinct values can be extracted (\cref{sec:synthetic-dataset,sec:qcm10-dataset,sec:iris-dataset}), to larger datasets (\cref{sec:wine-dataset}), where the exploration of general structure of the dataset is desirable.
In the following, \cref{fig:qcm,fig:iris-wine} provide a comparison of the quantitative data analysis of the PCP, RC, and our CPP with the \abbc{} and \abcd{} scheme. If not explicitly mentioned, linear scaling for all axes is employed.

\subsubsection{Synthetic Dataset}
\label{sec:synthetic-dataset}

Due to its simplicity, the dataset in \cref{fig:teaser} provides good access to briefly discuss interesting geometric properties of CPPs. Especially noteworthy is the cyan polygon in \cref{fig:teaser-cpp-abbc}, which constitutes a mirrored and translated variant of the orange polygon. This reflects reversal of the component order in $\datavec$ (which accounts for the mirroring along the main diagonal) and the addition of one to its entries (translating the polygon along the main diagonal).

\subsubsection{QCM10 Dataset}
\label{sec:qcm10-dataset}
With the QCM10 dataset~\cite{adak2020Classification} (\cref{fig:qcm}) from the UCI machine learning repository~\cite{dua2019}, we give an example of a typical application in the field of sensory measurement analysis. The dataset contains data of five gaseous alcohols (the five classes) with varying air-to-gas concentrations.
Due to the experimental design employed in the creation of this dataset, it contains an interesting property, in that all its members feature decreasing numeric values for increasing gas ratios (the five dimensions from first to last).

In the CPPs~(Figure~\ref{fig:qcm-abbc}--d), this trend is clearly visible, indicated by the majority of polygon vertices being located consistently below the main CPP diagonal, with edges connecting in counterclockwise order. This polygon shape signifies a negative correlation between adjacent value components (cf. \cref{sec:slopes-and-offsets}) and exemplifies the viability of our technique for trend identification inside (position of edges of a single polygon in image space) and across values (polygon edges in relation to each other).
In the PCP and RC, however, due to their independent normalization of each dimension, this correlation is hard to see, with some polylines featuring positive and some negative slopes.

\subsubsection{Iris Dataset}
\label{sec:iris-dataset}
This dataset~\cite{fisher1936Use} consists of 150 four-dimensional values containing physiological measurements about the iris flower, clustered into three classes (the three subspecies). It is a well researched and, in multi-dimensional visualization, often considered dataset. This popularity makes its CPP representation (Figure~\ref{fig:iris-abbc}--d) especially interesting, and promotes the comparability of our approach.

Apparent features standing out in the CPP \abcd{} scheme (\cref{fig:iris-abcd}) are the two-vertex polygons (i.e., lines). They signify an especially compact representation, where an entire four-dimensional value is tangible with a single line, which is generally not possible for the PCP and RC (\cref{fig:iris-pcp,fig:iris-rc}). More importantly, the \abcd{} scheme achieves full separation of the green cluster for this dataset, whereas the PCP and RC exhibit clutter similar to the \abbc{} CPP. This emphasizes the ability of our CPP with creation scheme \abcd{} to produce a compact representation that aides in cluster identification due to its cluttering reducing property; significantly more so than the correspondent PCP or RC representations.

\subsubsection{Wine Dataset}
\label{sec:wine-dataset}

The Wine dataset~\cite{forina1991parvus} from the UCI machine learning repository~\cite{dua2019}, features 178 13-dimensional values containing the chemical composition of wines, again clustered into three classes. Next to its clustering difficulty and general complexity for quantitative display, this dataset is especially worthy of consideration due to the differing number ranges between its dimensions, a property characteristic to its chemical content analysis.

\begin{figure}[t]
	\centering%
	\subfloat[\label{fig:iris-pcp}]{%
		\includegraphics[width=\fourImgWidth]{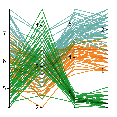}%
	}%
	\hfill%
	\subfloat[\label{fig:iris-rc}]{%
		\includegraphics[width=\fourImgWidth]{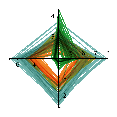}%
	}%
	\hfill%
	\subfloat[\label{fig:iris-abbc}]{%
		\includegraphics[width=\fourImgWidth]{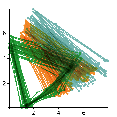}%
	}%
	\hfill%
	\subfloat[\label{fig:iris-abcd}]{%
		\includegraphics[width=\fourImgWidth]{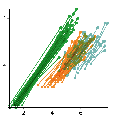}%
	}%
	\\
	\subfloat[\label{fig:wine-linear-pcp}]{%
		\includegraphics[width=\fourImgWidth]{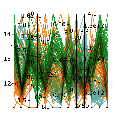}%
	}%
	\hfill%
	\subfloat[\label{fig:wine-linear-rc}]{%
		\includegraphics[width=\fourImgWidth]{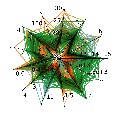}%
	}%
	\hfill%
	\subfloat[\label{fig:wine-linear-abbc}]{%
		\includegraphics[width=\fourImgWidth]{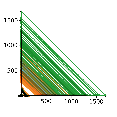}%
	}%
	\hfill%
	\subfloat[\label{fig:wine-linear-abcd}]{%
		\includegraphics[width=\fourImgWidth]{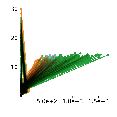}%
	}%
	\\
	\subfloat[\label{fig:wine-log-pcp}]{%
		\includegraphics[width=\fourImgWidth]{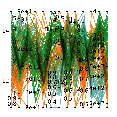}%
	}%
	\hfill%
	\subfloat[\label{fig:wine-log-rc}]{%
		\includegraphics[width=\fourImgWidth]{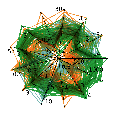}%
	}%
	\hfill%
	\subfloat[\label{fig:wine-log-abbc}]{%
		\includegraphics[width=\fourImgWidth]{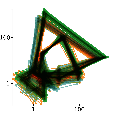}%
	}%
	\hfill%
	\subfloat[\label{fig:wine-log-abcd}]{%
		\includegraphics[width=\fourImgWidth]{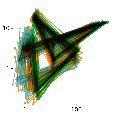}%
	}%
	\caption{\label{fig:iris-wine}
          Iris dataset~(4D, linear scale \protect\subref{fig:iris-pcp}--\protect\subref{fig:iris-abcd}) and 13-dimensional Wine dataset~(linear scale \protect\subref{fig:wine-linear-pcp}--\protect\subref{fig:wine-linear-abcd} and logarithmic scale \protect\subref{fig:wine-log-pcp}--\protect\subref{fig:wine-log-abcd}). Column arrangement analogous to \cref{fig:qcm}.
        }%
\end{figure}

Utilization of regular, linear scales (\cref{fig:wine-linear-pcp,fig:wine-linear-rc,fig:wine-linear-abbc,fig:wine-linear-abcd}) leads to crowded representations for the PCP, RC, \abbc{} CPP, and \abcd{} CPP. In particular, the CPPs are dominated by the large components in the data. 
However, employing logarithmic scaling on the plots (\cref{fig:wine-log-pcp,fig:wine-log-rc,fig:wine-log-abbc,fig:wine-log-abcd}) drastically increases readability and image-space utilization in the CPPs. Where a number of vertices in the linearly scaled CPPs lie close to the origin, they now contribute significantly to the resulting polygon shape, and uncover the principal relation of the components of the underlying $n$-dimensional value. Due to individually scaled dimensions, logarithmic scales have a noticeably lower impact with both the PCP and RC. This signifies an advantage of the single 2D space used by our technique, since logarithmic scaling not only uncovers additional structure but also simplifies interpretation of the plot due to the single pair of axes.

\subsection{Placement}
\label{sec:results-placement}


\begin{figure}[t]
	\captionsetup[subfloat]{textfont=tiny}%
	\centering%
	\subfloat[\label{fig:iris-placement-tsne}~$\ji=\textbf{0.95}$, $\Sc=0.63$]{%
		\includegraphics[width=\threeImgWidth]{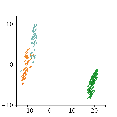}%
	}%
	\hfill%
	\subfloat[\label{fig:iris-placement-umap}~$\ji=0.93$, $\Sc=0.64$]{%
		\includegraphics[width=\threeImgWidth]{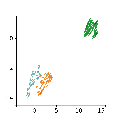}%
	}%
	\hfill%
	\subfloat[\label{fig:iris-placement-avg}~$\ji=0.86$, $\Sc=0.45$]{%
		\includegraphics[width=\threeImgWidth]{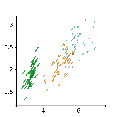}%
	}%
	\\
	\subfloat[\label{fig:iris-placement-geo}~$\ji=0.92$, $\Sc=0.57$]{%
		\includegraphics[width=\threeImgWidth]{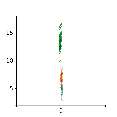}%
	}%
	\hfill%
	\subfloat[\label{fig:iris-placement-geo-abbc}~$\ji=0.69$, $\Sc=0.21$]{%
		\includegraphics[width=\threeImgWidth]{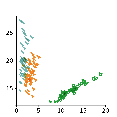}%
	}%
	\hfill%
	\subfloat[\label{fig:iris-placement-stat}~$\ji=0.85$, $\Sc=0.43$]{%
		\includegraphics[width=\threeImgWidth]{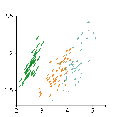}%
	}%
	\caption{\label{fig:iris-placement}
          Iris dataset (4D). Comparison of \mbox{t-SNE}~\protect\subref{fig:iris-placement-tsne} and UMAP~\protect\subref{fig:iris-placement-umap} with our intrinsic~\protect\subref{fig:iris-placement-avg}, geometric (\abcd{})~\protect\subref{fig:iris-placement-geo}, geometric (\abbc{})~\protect\subref{fig:iris-placement-geo-abbc}, and statistical~\protect\subref{fig:iris-placement-stat} placement. Subcaptions refer to the corresponding Jaccard index ($\ji$) and silhouette coefficient ($\Sc$). Bold values denote best performance (see also \cref{tab:clustering}).
        }%
\end{figure}

Before discussing the qualitative comparison of our placement strategies, we derive a quantitative measure to evaluate placements. To do this effectively, we evaluate the k-means partitioning of a placement versus the true classification labels of the corresponding dataset, by calculating the Jaccard index~\cite{jaccard1912Distribution}~($\ji \in [0,1]$, higher is better) as a measure of similarity between two classifications and the silhouette coefficient~\cite{rousseeuw1987Silhouettes}~($\Sc \in [-1,1]$, higher is better) as a measure of value--cluster proximity.

We compare (\cref{tab:clustering}) all four variants of our placement (\cref{sec:placement}) using the \abbc{} and \abcd{} scheme to \mbox{t-SNE} (perplexity = 5, 30, 80) and UMAP (nNeighbors = 5, 15, 50) (intrinsic placement is not calculated for the \abbc{} scheme, statistical placement is invariant to both schemes).
This serves as a baseline for the following discussion, where we present some detailed examples of our placement. If not explicitly mentioned, the following \cref{fig:iris-placement,fig:wine-placement,fig:billiard-placement} depict t-SNE and UMAP (in their best configuration according to \cref{tab:clustering}), and our four placements using the \abcd{} scheme.

\subsubsection{Iris Dataset}
\label{sec:iris}

Whereas the CPP with the \abcd{} scheme proved especially strong in separating the green cluster (\cref{fig:iris-abcd}), the geometric placement performance of \abcd{} in \cref{fig:iris-placement-geo}  ($\ji=0.92$, $\Sc=0.574$) is competitive with the best clustering result (\cref{fig:iris-placement-tsne}, $\ji=0.945$, $\Sc=0.632$)

Especially still, our geometric \abbc{} scheme placement (\cref{fig:iris-placement-geo-abbc}, $\ji=0.708$, $\Sc=0.21$) (which we visualize instead of the, in the two-vertex polygon case insignificant, angular placement) profits from our scaled-down polygons (compared to simple points), since it enables the separation of the orange and cyan cluster by comparing the different wedge shape between the polygons of each cluster.

Interesting to note is the similar performance of the statistical placement (\cref{fig:iris-placement-stat}, $\ji=0.853$, $\Sc=0.429$) to t-SNE and UMAP (\cref{fig:iris-placement-tsne,fig:iris-placement-umap}), as this relation also carries significance for the following considered datasets (\cref{sec:wine,sec:billiard}).

\subsubsection{Wine Dataset}
\label{sec:wine}

\begin{figure}[t]
	\captionsetup[subfloat]{textfont=tiny}%
	\centering%
	\subfloat[\label{fig:wine-placement-tsne}~$\ji=0.73$, $\Sc=0.26$]{%
		\includegraphics[width=\threeImgWidth]{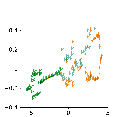}%
	}%
	\hfill%
	\subfloat[\label{fig:wine-placement-umap}~$\ji=0.72$, $\Sc=0.25$]{%
		\includegraphics[width=\threeImgWidth]{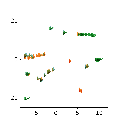}%
	}%
	\hfill%
	\subfloat[\label{fig:wine-placement-avg}~$\ji=\textbf{0.80}$, $\Sc=0.30$]{%
		\includegraphics[width=\threeImgWidth]{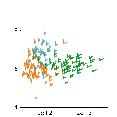}%
	}%
	\\
	\subfloat[\label{fig:wine-placement-geo}~$\ji=0.58$, $\Sc=0.10$]{%
		\includegraphics[width=\threeImgWidth]{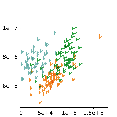}%
	}%
	\hfill%
	\subfloat[\label{fig:wine-placement-ang}~$\ji=0.70$, $\Sc=0.08$]{%
		\includegraphics[width=\threeImgWidth]{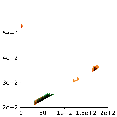}%
	}%
	\hfill%
	\subfloat[\label{fig:wine-placement-stat}~$\ji=0.70$, $\Sc=0.20$]{%
		\includegraphics[width=\threeImgWidth]{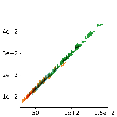}%
	}%
	\caption{\label{fig:wine-placement}
          Wine dataset (13-dimensional). Comparison of \mbox{t-SNE}~\protect\subref{fig:wine-placement-tsne}, UMAP~\protect\subref{fig:wine-placement-umap}, and our intrinsic~\protect\subref{fig:wine-placement-avg}, geometric~\protect\subref{fig:wine-placement-geo}, angular~\protect\subref{fig:wine-placement-ang}, and statistical~\protect\subref{fig:wine-placement-stat} placement.
        }%
\end{figure}

Here, our intrinsic placement (\cref{fig:wine-placement-avg}, $\ji=0.803$, $\Sc=0.296$) constitutes the best cluster separation compared to t-SNE (\cref{fig:wine-placement-tsne}, $\ji=0.725$, $\Sc=0.259$) and UMAP (\cref{fig:wine-placement-umap}, $\ji=0.724$, $\Sc=0.25$).
Additionally, our small polygons feature distinct shapes which improves individual value identification in the placement.

Furthermore, notice that the relatively worse performance of the statistical placement (\cref{fig:wine-placement-stat}, $\ji=0.702$, $\Sc=0.201$) is again analogous to UMAP in terms of $\ji$ and $\Sc$, and also w.r.t.\ qualitative expressiveness. This suggests a correlation between measures directly calculated from the $n$-dimensional value and dimensionality reduction results, confirming that the utilization of geometric polygon properties for placement is justified and beneficial.

\subsubsection{Billiard Dataset}
\label{sec:billiard}

Generally, phase spaces of billiard dynamics~\cite{tabachnikov2005Geometry} are visualized in a 2D plot, where two orthogonal axes are used to display its two components, i.e., angle and arclength. This works great for single trajectories, but is unsuited for comparing sets of trajectories.

To test the suitability of our approach for this type of analysis, we employ it to the Billiard dataset, containing the phase space of 60 2D elliptical billiard dynamics~\cite{tabachnikov2005Geometry} trajectories with 50 reflections each. The starting position and direction of each trajectory are seeded three degrees apart from its predecessor, and the trajectories are split into three clusters by varying the parameter~$A$ of the elliptical border ${x^2}/{A^2}+{y^2}/{B^2}=1$ by $0.1$ each. 

\begin{figure}[t]%
	\captionsetup[subfloat]{textfont=tiny}%
	\centering%
	\subfloat[\label{fig:billiard-placement-tsne}$\ji=0.79$, $\Sc=0.32$]{%
		\includegraphics[width=\threeImgWidth]{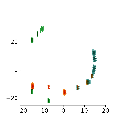}%
	}%
	\hfill%
	\subfloat[\label{fig:billiard-placement-umap}$\ji=0.72$, $\Sc=0.25$]{%
		\includegraphics[width=\threeImgWidth]{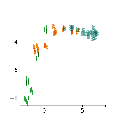}%
	}%
	\hfill%
	\subfloat[\label{fig:billiard-placement-avg}$\ji=0.70$, $\Sc=0.38$]{%
		\includegraphics[width=\threeImgWidth]{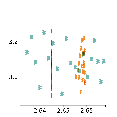}%
	}%
	\\
	\subfloat[\label{fig:billiard-placement-geo}$\ji=\textbf{1.00}$, $\Sc=0.98$]{%
		\includegraphics[width=\threeImgWidth]{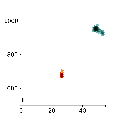}%
	}%
	\hfill%
	\subfloat[\label{fig:billiard-placement-ang}$\ji=\textbf{1.00}$, $\Sc=0.66$]{%
		\includegraphics[width=\threeImgWidth]{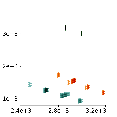}%
	}%
	\hfill%
	\subfloat[\label{fig:billiard-placement-stat}$\ji=0.35$, $\Sc={-0.03}$]{%
		\includegraphics[width=\threeImgWidth]{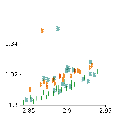}%
	}%
	\caption{\label{fig:billiard-placement}
          Billiard dataset (100-dimensional). Comparison of \mbox{t-SNE}~\protect\subref{fig:wine-placement-tsne} and UMAP~\protect\subref{fig:wine-placement-umap} with our intrinsic~\protect\subref{fig:wine-placement-avg}, geometric~\protect\subref{fig:wine-placement-geo}, angular~\protect\subref{fig:wine-placement-ang}, and statistical~\protect\subref{fig:wine-placement-stat} placement.
        }%
\end{figure}

Comparing clustering performance alone, the geometric placement (\cref{fig:billiard-placement-geo}, $\ji=1$, $\Sc=0.977$) delivers a nearly perfect result, with complete separation of all three clusters. However, the angular placement (\cref{fig:billiard-placement-ang}, $\ji=1$, $\Sc=0.657$) shows a wider distribution of the values in image space, encouraging leveraging of this additional information in qualitative analysis. Compared to the previous datasets, the characteristic of billiard dynamics phase space seems to suit the angular placement especially well, possibly because a significant amount of inherent information is related to angles.

Crucially, the statistical placement (\cref{fig:billiard-placement-stat}, $\ji=0.35$, $\Sc=-0.033$) fails to provide a clearly separated embedding of the three clusters, visually in line with both t-SNE (\cref{fig:billiard-placement-tsne}, $\ji=0.785$, $\Sc=0.324$) and UMAP (\cref{fig:billiard-placement-umap}, $\ji=0.74$, $\Sc=0.337$) results, further reinforcing our theory of their analogy in terms of clustering performance discussed in \cref{sec:iris}.

\begin{table*}[b]
\centering
\caption{\label{tab:clustering}
  Clustering performance comparing CPP placement strategies with the 2D embedding of \mbox{t-SNE} and UMAP, in terms of Jaccard index~$\ji$ and silhouette coefficient~$\Sc$. The \mbox{t-SNE} and UMAP results refer to the average result after ten runs each, to compensate for possible variance between single runs. Bold values denote best performance per column for the Jaccard index. Best performance of the silhouette coefficient is not emphasized, since interpretation of this value holds little significance without a corresponding Jaccard similarity value.
}%
\label{tab:clustering}
\begin{tabularx}{\textwidth}{@{}Xr r rrr rr rrr rrr@{}} 
\toprule
Approach &    & \multicolumn{6}{c}{Cyclic Polygon Plot} & \multicolumn{3}{c}{t-SNE} & \multicolumn{3}{c}{UMAP} \\
\cmidrule(lr){3-8}
\cmidrule(lr){9-11}
\cmidrule(lr){12-14} 
Configuration &    &   & \multicolumn{3}{c}{\abcd{}} & \multicolumn{2}{c}{\abbc{}} & \multicolumn{3}{c}{perplexity} & \multicolumn{3}{c}{nNeighbors} \\
\cmidrule(lr){4-6}
\cmidrule(lr){7-8}
\cmidrule(lr){9-11}
\cmidrule(lr){12-14}
Placement &    & \multicolumn{1}{c}{stat} & \multicolumn{1}{c}{int} & \multicolumn{1}{c}{geo} & \multicolumn{1}{c}{ang} & \multicolumn{1}{c}{geo} & \multicolumn{1}{c}{ang} & \multicolumn{1}{c}{5} & \multicolumn{1}{c}{30} & \multicolumn{1}{c}{80} & \multicolumn{1}{c}{5} & \multicolumn{1}{c}{15} & \multicolumn{1}{c}{50}  \\ 
\midrule
\multirow{2}{*}{Iris} & $\Sc$ & 0.429 & 0.448 & 0.574 & \multicolumn{1}{c}{—} & 0.213 & 0.649 & 0.562 & 0.632 & 0.589 & 0.597 & 0.649 & 0.638 \\
& $\ji$ & 0.853 & 0.860 & 0.920 & \multicolumn{1}{c}{—} & 0.693 & 0.633 & 0.943 & \textbf{0.945} & 0.927 & 0.926 & 0.905 & 0.934 \\
\multirow{2}{*}{Billiard} & $\Sc$ & -0.033 & 0.379 & 0.977 & 0.657 & -0.015 & -0.113 & 0.324 & 0.312 & -0.049 & 0.177 & 0.337 & 0.253 \\
& $\ji$ & 0.350 & 0.700 & \textbf{1.000} & \textbf{1.000} & 0.450 & 0.367 & 0.785 & 0.735 & 0.420 & 0.567 & 0.740 & 0.740 \\
\multirow{2}{*}{Wine} & $\Sc$ & 0.201 & 0.296 & 0.104 & 0.076 & 0.210 & 0.053 & 0.258 & 0.258 & 0.259 & 0.187 & 0.187 & 0.250 \\
& $\ji$ & 0.702 & \textbf{0.803} & 0.584 & 0.697 & 0.708 & 0.517 & 0.724 & 0.724 & 0.725 & 0.653 & 0.653 & 0.724 \\
\bottomrule
\end{tabularx}
\end{table*}

\subsection{User Study}
\label{sec:results-userstudy}

In order to provide a quantitative assessment on the properties of our approach, we conducted a user study comparing the CPP to the PCP and RC. To achieve quantitative and comparable results, the user study was focused on the CPP without placement, since its evaluation we already discussed above. Additionally, the CPP was employed without start arrows and vertex circles, to limit the scope of the study to tasks that do not rely on this information (\cref{sec:results-userstudy-tasks}) and improve comparability to the other techniques. A representative sample of images used in the study for all tasks is available in the supplemental material together with further detail.

\subsubsection{Tasks}
\label{sec:results-userstudy-tasks}

We focus on measuring user performance~\cite{wehrend1990problemoriented} by comparing the approaches \emph{head to head}~\cite{lam2012Empirical,isenberg2013Systematic} in three main analytic tasks, which are widely employed in multi-dimensional data analysis~\cite{amar2005Lowlevel}:
\begin{itemize}[noitemsep,topsep=-6pt]
	\item outlier detection (OD),
	\item value retrieval (VR), and
	\item value comparison (VC).
          \vspace*{6pt}
\end{itemize}

\noindent Formulation of these tasks motivated the following hypotheses.

\subsubsection{Hypotheses}
\paragraph*{H1.}
We assume that the CPP will see less of a degradation in task accuracy when moving from five- to ten-dimensional data than the PCP, due to PCP's innate axis-cluttering when displaying higher-dimensional data. 

\paragraph*{H2.}
We assume that the value retrieval task will perform better with CPPs than RCs, because, even though both being polygon based visualizations, the cyclic polygon plot benefits from a simple 2D space. This should especially hold true for the outlier detection task in comparison to both other approaches (PCP and RC). 

\paragraph*{H3.}
When comparing the two creation schemes, we assume that \abbc{} will perform better in terms of task accuracy and completion time for lower-dimensional data (more structure) and \abcd{} better for higher-dimensional
(less overdrawing).

\paragraph*{}
We expect H1 to hold true to at least similar extent when comparing to the RC instead of the PCP, due to its inferiority in this regard to linear layouts~\cite{waldner2020Comparison,goldberg2011Eye}.

\subsubsection{Datasets}

To effectively measure the performance of the three tasks, we created specific datasets for each of the tasks. All datasets were created with up to ten ten-dimensional values. We chose the maximum number of ten members per dataset to be able to support ten uniquely colored polylines/polygons. For this, we used a \textit{10-color-paired} color scheme with light and dark shades of five different color hues. 

The first dataset was created by filling all components of all $n$-dimensional values with random uniform noise in the interval $[0,0.8]$, and then inserting a single random number in the interval $[0.8,1]$. This dataset was used for the outlier detection task. 

The second dataset was created by first inserting random uniform noise in the interval $[0,1]$ in all components of the $n$-dimensional value. Then a single, manually defined numeric value, was inserted as one of the components in a random dataset member. This dataset was used for the outlier detection and value retrieval tasks. 

The third dataset was created analogous to the second dataset, with the exception of inserting two but one numeric values in the same fashion. This dataset was used for the value comparison task. In order to also account for datasets with different number ranges per dimension, which is frequently the case for datasets containing sensory measurements, for dataset type two and three, random scaling factors for each of the dimensions were employed, to scale all numeric values accordingly. These factors were used for half of the visualizations displayed in the study, the other half used datasets with values in the $[0,1]$ range.

\subsubsection{Questions}
Derived from the tasks in \cref{sec:results-userstudy-tasks}, the following questions were designed to accompany the visualizations of the previously discussed datasets in the study. For the outlier detection task, two questions were designed, which were used alternately:
\paragraph*{OD}
\begin{itemize}[noitemsep,topsep=-6pt]
	\item \textit{Is there an attribute value greater 0.8 present in the dataset?}
	\item \textit{Select the color of the polyline / radar glyph / cyclic polygon representing the data-vector with the largest attribute value in the dataset.}
\end{itemize}
\paragraph*{VR}
\begin{itemize}[noitemsep,topsep=-1pt]
	\item \textit{Does the displayed dataset contain an attribute value of exactly $X$?}
\end{itemize}
Varying phrasing of the value comparison question was necessary to suit the three different techniques:
\paragraph*{VC}
\begin{itemize}[noitemsep,topsep=-6pt]
	\item PCP: \textit{Is the attribute value represented by the indicated axis-intercept of polyline A or polyline B larger?}
	\item RC: \textit{Is the attribute value represented by the indicated axis-intercept of radar glyph A or radar glyph B larger?}	
	\item CPP: \textit{Is the attribute value represented by the $x$/$y$-coordinate of vertex A or the $x$/$y$-coordinate of vertex B larger?}	
\end{itemize}
\vspace{6pt}

\begin{figure*}
	\captionsetup[subfloat]{captionskip=-3pt}%
	\centering%
	\subfloat[\label{fig:userstudy-pOD5} OD, 5D]{%
		\includegraphics[width=\sixImgWidth]{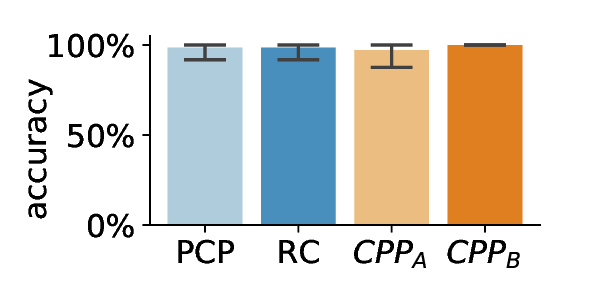}%
	}%
	\hfill%
	\subfloat[\label{fig:userstudy-pOD10} OD, 10D]{%
		\includegraphics[width=\sixImgWidth]{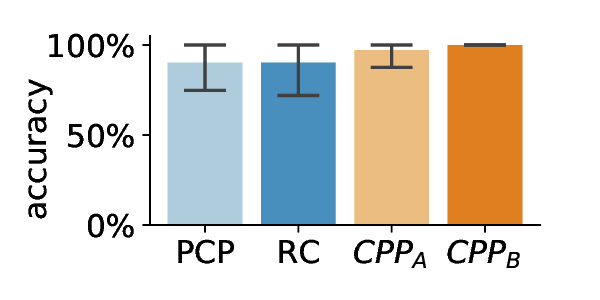}%
	}%
	\hfill%
	\subfloat[\label{fig:userstudy-pVR5} VR, 5D]{%
		\includegraphics[width=\sixImgWidth]{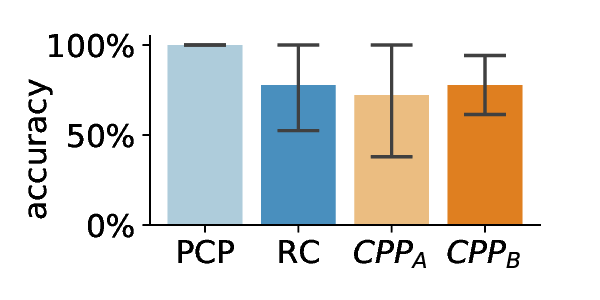}%
	}%
	\hfill%
	\subfloat[\label{fig:userstudy-pVR10} VR, 10D]{%
		\includegraphics[width=\sixImgWidth]{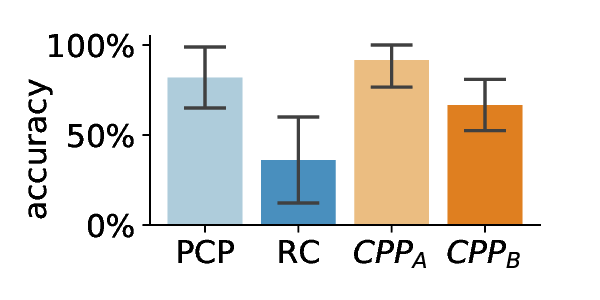}%
	}%
	\hfill%
	\subfloat[\label{fig:userstudy-pVC5} VC, 5D]{%
		\includegraphics[width=\sixImgWidth]{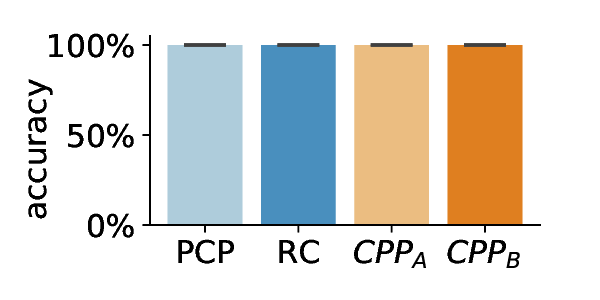}%
	}%
	\hfill%
	\subfloat[\label{fig:userstudy-pVC10} VC, 10D]{%
		\includegraphics[width=\sixImgWidth]{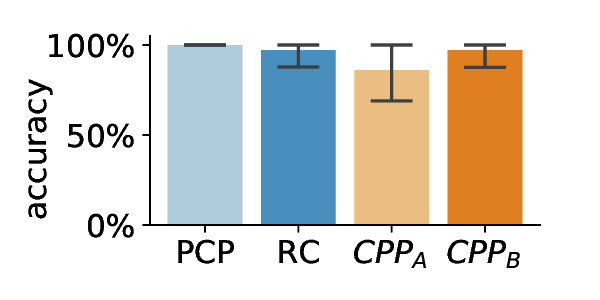}%
	}%
	\\
	\captionsetup[subfloat]{captionskip=-6pt}%
	\subfloat[\label{fig:userstudy-tOD5} OD, 5D]{%
		\includegraphics[width=\sixImgWidth]{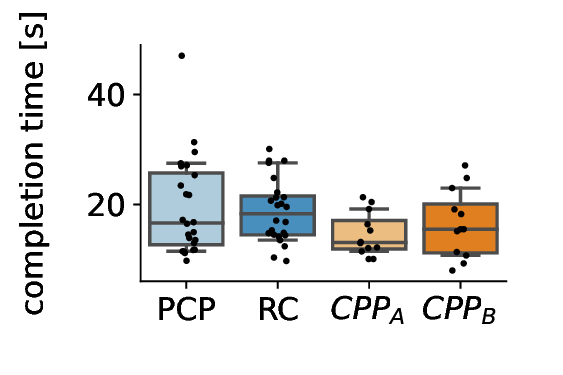}%
	}%
	\hfill%
	\subfloat[\label{fig:userstudy-tOD10} OD, 10D]{%
		\includegraphics[width=\sixImgWidth]{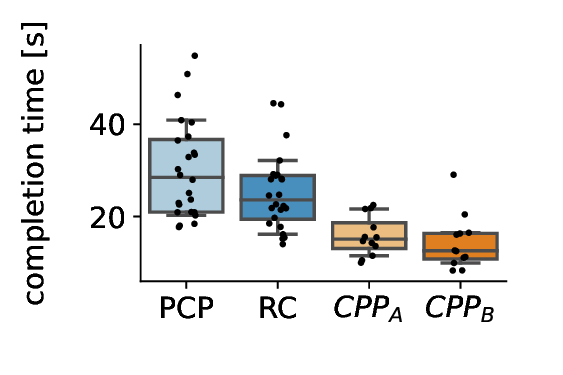}%
	}%
	\hfill%
	\subfloat[\label{fig:userstudy-tVR5} VR, 5D]{%
		\includegraphics[width=\sixImgWidth]{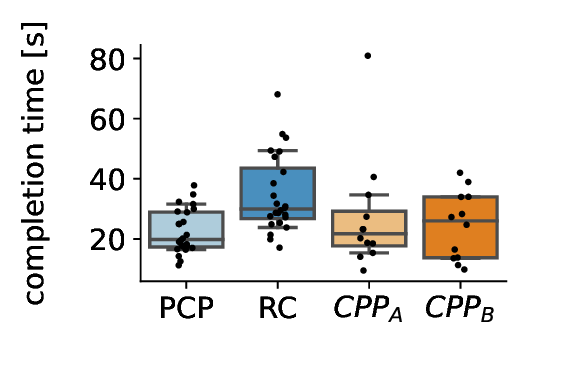}%
	}%
	\hfill%
	\subfloat[\label{fig:userstudy-tVR10} VR, 10D]{%
		\includegraphics[width=\sixImgWidth]{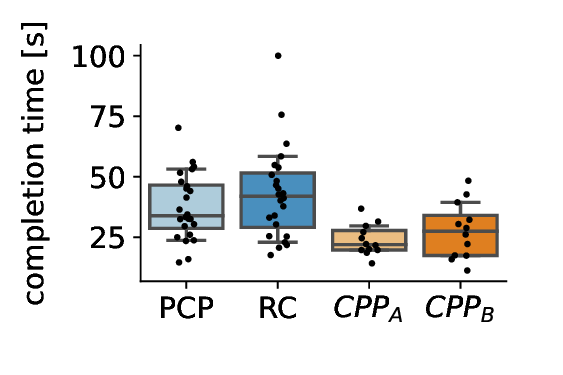}%
	}%
	\hfill%
	\subfloat[\label{fig:userstudy-tVC5} VC, 5D]{%
		\includegraphics[width=\sixImgWidth]{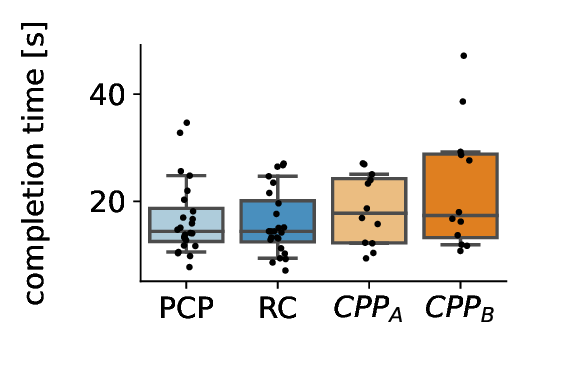}%
	}%
	\hfill%
	\subfloat[\label{fig:userstudy-tVC10} VC, 10D]{%
		\includegraphics[width=\sixImgWidth]{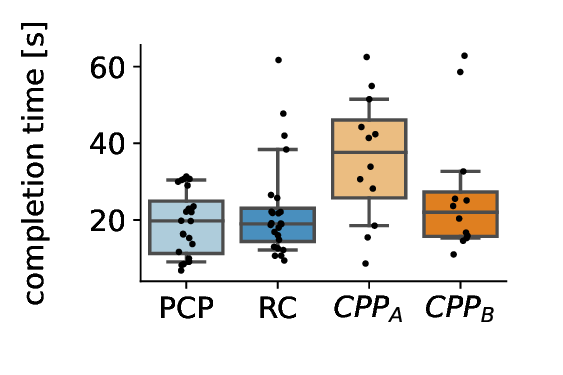}%
	}%
	\caption{\label{fig:userstudy}
          Results of the user study. First row shows task accuracy in percent for outlier detection~(OD), value retrieval~(VR), and value comparison~(VC) for the five-~(5D) and ten-dimensional~(10D) data. Second row analogously shows task completion time. Whiskers in the bar plot show the confidence interval of one standard deviation. They are upper bound to 100\% in the case where their value would exceed 100\%. Whiskers of the box plot represent the 10--90 percentile interval around the median, the box itself represents the interquartile range (Q1--Q3). CPP$_A$ denotes the cyclic polygon plot with \abcd{} creation type, CPP$_B$ the cyclic polygon plot with \abbc{} creation type.
        }%
\end{figure*}

\subsubsection{Design}

Due to the COVID-19 pandemic, we held the user study online with 24 participants. We recruited them from the university environment, aging 22 to 55 years. All participants were given a detailed and live introduction and presentation via video call of 30--40 minutes on multi-dimensional data analysis, and more specifically, the three employed techniques. The introductory videos presented to the participants are provided in the supplemental material. Three example questions (one for each technique) were solved together, with feedback provided.

We used a Likert scale to have participants rate their experience in multi-dimensional data analysis from 1~(no experience) to 5~(expert). The mean value over all participants was 2.5 with three people selecting one and only one person selecting five as their experience level. Additionally, participants were presented the color scheme and asked to confirm their ability to discern all the displayed colors. 

The study itself consisted of 54 questions, categorized as follows. For all three tasks, five- and ten-dimensional datasets were used with three questions per technique, per task, and per dataset dimensionality. Versions A and B of the study were created, where version A contained the CPP with the \abcd{} scheme and version B contained the CPP with the \abbc{} scheme. All other questions remained exactly the same. The questions were shown in the order PCP, RC, CPP while the order of questions per task was randomized to minimize learning effects across the study duration. One half of participants were shown study A, the other half was shown study B. A representative sample of visualizations used for the study is provided in the supplemental material.

\begin{table}[b]
  \caption{\label{tab:userstudy-time-ANOVA}
    ANOVA results for the completion time of the user study for outlier detection (OD), value retrieval (VR), and value comparison (VC). 5D/10D denote the five- and ten-dimensional dataset.%
  }%
	\centering
	\begin{tabularx}{\linewidth}{X  rr}
		\toprule
		\multicolumn{1}{l}{Task} & \multicolumn{1}{c}{F-value} & \multicolumn{1}{c}{p-value} \\
		\midrule
		OD, 5D	&1.737&$1.68\times 10^{-1}$\\
		OD, 10D	&13.606&$4.76\times 10^{-7}$\\
		VR, 5D	&4.231&$8\times 10^{-3}$\\
		VR, 10D	&5.957&$1\times 10^{-3}$\\
		VC, 5D	&2.208&$9.5\times 10^{-2}$\\
		VC, 10D	&4.772&$4\times 10^{-3}$\\
		\bottomrule
	\end{tabularx}
\end{table}

\subsubsection{Study Results}
\label{sec:results-userstudy-results}

We first determined the statistical significance of our completion time results with ANOVA ($\alpha = 0.05$). The critical F-value for our study setup for all tasks is $F(3,68)=2.740$. The five-dimensional outlier detection and value comparison tasks have an F-value below this critical threshold, rendering them statistically not significant. For all other tasks, a statistically significant difference in variance between the visualizations for our chosen $\alpha$ exists (\cref{tab:userstudy-time-ANOVA}). 

No ANOVA was performed on the results of the task accuracy, since due to the design of our study (categorical, single choice answers), the assumption of normality is not given and ANOVA would not yield significant results~\cite{jaeger2008Categorical}. Additionally, we argue that due to the narrow scope of the study, task accuracy detailed in \cref{fig:userstudy-pOD5,fig:userstudy-pOD10,fig:userstudy-pVR5,fig:userstudy-pVR10,fig:userstudy-pVC5,fig:userstudy-pVC10} provides sufficient expressiveness over the performance of the approaches.

The results of our study (\cref{fig:userstudy}) show competitive properties of the CPP, especially promising in the ten-dimensional setting, as well as a sizeable advantage regarding the completion time across all settings of the outlier detection and value retrieval tasks. These results are detailed by the following discussion per task.

\subsubsection{Outlier Detection}
The five-dimensional setting~(\cref{fig:userstudy-pOD5,fig:userstudy-tOD5}) was managed well by all four tested approaches. CPPs with the \abcd{} scheme performed slightly worse than their \abbc{} counterpart, which could be attributed to the redundancy present in the \abbc{} configuration. While scanning the visualization for outliers in the \abbc{} configuration, it is sufficient to focus either on the horizontal or the vertical axis.

In the ten-dimensional context, we see a more evident difference in accuracy between the approaches~(\cref{fig:userstudy-pOD10}). Both CPPs exhibit higher accuracy than the PCP and RC visualizations, with the \abbc{} scheme again performing even slightly better (H1). Additionally, completion time in the ten-dimensional setting is especially noteworthy, since here, CPPs performed on average almost two times faster than RCs and more than two times faster than PCPs (\cref{fig:userstudy-tOD10}).

\subsubsection{Value Retrieval}
In the five-dimensional context, the PCP clearly performed best in both accuracy and completion time~(\cref{fig:userstudy-pVR5,fig:userstudy-tVR5}). This can be attributed to the manageability of five dimensions by the PCP, which makes comparison between axes easy and fast. RCs exhibit the worst completion time for both five- and ten-dimensional settings, while also performing comparably in terms of accuracy in the five-dimensional setting compared to the CPP approaches.

Again, the move to ten-dimensional data shows a notable increase in performance for our CPPs~(\cref{fig:userstudy-pVR10,fig:userstudy-tVR10}). Whereas PCP (lower accuracy and higher completion time) and RC (lower accuracy with equal completion time) show a decline in their performance compared to the five-dimensional setting, the CPP visualizations actually improve on both task accuracy and completion time. This confirms our previously discussed hypothesis (H1), that the CPP can adapt better to higher-dimensional data, and additionally shows that the lack of individual axes is not critical to the interpretability of our plot. In this setting, the CPP with the \abcd{} scheme shows the most significant increase in accuracy compared to the other approaches, confirming H2.

\subsubsection{Value Comparison}
Both RC and PCP performed comparably well in both accuracy and completion time~(\cref{fig:userstudy-pVC5,fig:userstudy-tVC5}). Since in this configuration, values to compare were already highlighted in the visualization, the viewer could focus on the significant visualization parts of the RC and PCP, which reduced their complexity in the ten-dimensional setting. 

While performing equally well in the five-dimensional context, the CPP showed a deterioration of accuracy of about 10\% (\abbc{}) and 20\% (\abcd{})~(\cref{fig:userstudy-pVC10}). More crucially, task completion time for the CPPs were especially slow in the value comparison task~(\cref{fig:userstudy-tVC10}). 
This uncovers a drawback of the CPP which makes it necessary to refer to $x$- and $y$-components of the vertices to unambiguously refer to a single attribute value, which can be attributed to the necessary lack of individual axis labels in favor of screen-space efficiency. This aggravates the task description and required additional time for the participants to interpret the question which is evident from the longer completion time. Additionally, in the 10-dimensional context, this manifests itself in a lower task accuracy, which could be attributed to the complexity of the plot, which already exhibited some amount of overdrawing.

H3 only held true in context of value retrieval. For both other tasks, the \abbc{} had slight accuracy advantage in all configurations.

\section{Discussion and Limitations}
\label{sec:disc}
We have investigated the CPP with two cyclic selection schemes and four variants of its placement. Our results have shown that the \abcd{} scheme provides a good baseline for all discussed datasets, resulting in a valid and expressive visualization. In lower-dimensional datasets especially, we suggest the \abbc{} scheme to uncover additional dataset structure, which our results confirm.

Regarding placement of the polygons, the intrinsic placement strategy proved as a solid baseline, showing competitive performance for all datasets discussed in our results.
Employment of the angular placement is mostly limited to niche applications, but can be especially advantageous when used with suitable datasets, e.g., the Billiard dataset.
Statistical placement shows little significance beyond the fact that it confirms the validity of our other placements, which are motivated directly from the polygon geometry and generally show better performance.
Motivated by these results, we recommend the use of intrinsic placement as the default placement strategy as it provides very competitive results for all datasets and suggest our other placement types as supplemental and application-specific. Additionally, we recommend our placement over other optimization-based approaches like t-SNE and UMAP, since it preserves a strong correspondence to our polygons, which, when viewed as small glyphs, convey additional information about the underlying data.

The difficulty in representing identical vertices of a polygon, which we addressed with compositing in the rendering step, is still an innate drawback, but, as our results and user study show, has little impact in practice when applying the CPP to real datasets.

The comparatively worse performance of the CPP in the value comparison task of the user study emphasizes another innate drawback of the CPP in its complexity of referring to explicit components of a polygon vertex. Whereas specific value components in techniques featuring separate coordinate axes per dimension can straightforwardly be referred to, the CPP necessitates closer inspection of a polygon with its starting arrow. While this circumstance can in part be attributed to the lack of individual axis labels, necessary for our screen-space efficient design, the user study and the application of our technique to real datasets again show that it is nevertheless competitively performant for key visualization tasks.

Finally, as it is a shared drawback of line-based approaches, cluttering and overdrawing still remains present in CPPs of higher dimensional data. We have shown, however, that, dependent on the displayed data, this problem can be alleviated in the CPP by using logarithmic scaling on the axes, which is especially effective in decompressing previously crowded areas in our plot, as we discussed in \cref{sec:wine-dataset}.

\section{Conclusion}
\label{sec:conc}

We introduced the cyclic polygon plot, a novel approach to visualize $n$-dimensional discrete data, based on decomposition of the original $n$D value into 2D subspaces, whose 2D points are projected to image space. A polygon representation preserves correspondence to the original data dimensions. We conducted a detailed evaluation and discussion of its properties, backed up by a a user study. Additionally, we derived glyphs from our approach, and presented novel strategies to place these glyphs based on their intrinsic properties, resulting in an approach that we compare to existing dimensionality reduction techniques. Although our approach outperforms existing techniques in some cases, it also exhibits limitations, including difficulties with identical values. Future work could research alternative representation of such multiple values.

\acknowledgments{
This work is supported by the Deutsche Forschungsgemeinschaft (DFG, German Research Foundation) – Project-ID 281071066 – TRR 191 (Transregional Colloborative Research Center SFB / TRR 191) and Germany’s Excellence Strategy EXC2181/1 - 390900948 (the Heidelberg STRUCTURES Excellence Cluster).}

\bibliographystyle{abbrv}

\bibliography{literature.bib}
\end{document}